\documentclass[aps,amsfonts,pra,twocolumn,showpacs,superscriptaddress]{revtex4-1}
\usepackage{epsfig,amsmath,amssymb,bm,epsf,graphicx,psfrag,float}

\usepackage{amsfonts}

\usepackage{color}
\usepackage{xcolor}
\usepackage{mathtools}

\usepackage{bbm}
\usepackage{enumerate}
\usepackage{subfigure}

\usepackage[all]{xy}
\usepackage{color}
\usepackage[utf8]{inputenc}
\usepackage[T1]{fontenc}
\usepackage{lmodern, mathtools}
\newtheorem{prop}{Proposition}
\usepackage{tikz}
\usetikzlibrary{shapes,positioning,arrows,calc,intersections,through,patterns}

\newcommand{\op}[1]{\operatorname{#1}}

\begin{document}
\tikzstyle{Lv}=[circle,fill=black!70!blue,inner sep=0pt,minimum size=1mm]
\tikzstyle{Cv}=[circle,fill=black!70!red,inner sep=0pt,minimum size=1mm]
\tikzstyle{Rv}=[circle,fill=black!70!green,inner sep=0pt,minimum size=1mm]

\title{Demonstrating the AKLT spectral gap on 2D degree-3 lattices}

\author{Nicholas Pomata}
\affiliation{C. N. Yang Institute for Theoretical Physics and Department of Physics and Astronomy, State University of New York at Stony Brook, Stony Brook, NY 11794-3840, USA}
\author{Tzu-Chieh Wei}
\affiliation{C. N. Yang Institute for Theoretical Physics and Department of Physics and Astronomy, State University of New York at Stony Brook, Stony Brook, NY 11794-3840, USA}
\affiliation{Institute for Advanced Compuational Science, State University of New York at Stony Brook, Stony Brook, NY 11794-5250, USA}
\date{\today}

\begin{abstract}
We establish that the spin-3/2 AKLT model on the honeycomb lattice has a nonzero
spectral gap. We use the relation between the anticommutator of two projectors
and their sum, and apply it to related AKLT projectors that occupy plaquettes 
or other extended regions. We analytically reduce the complexity in
the resulting eigenvalue problem and use a Lanczos numerical method to show that
the required inequality for the nonzero spectral gap holds. This approach is
also successfuly applied to several other spin-3/2 AKLT models on degree-3
semiregular tilings, such as the square-octagon, star and cross lattices, where
the complexity is low enough that exact diagonalization can be used instead
of the Lanczos method. In addition, we also close the previously open cases in
the singly decorated honeycomb and square lattices.
\end{abstract}

\maketitle

\section{Introduction}

The question of the existence of a spectral gap above the lowest energy state(s)
plays a key factor in topological phases of matter and many properties
of interacting systems~\cite{Wen1989,Gu2009,Pollmann2012}.
It is fundamentally related to the
Lieb-Schultz-Mattis theorem~\cite{LSM} and generalizations
thereof~\cite{HastingsLSM,Oshikawa} as well as the Haldane
gapped phases~\cite{Haldane1D}, where there is a symmetry in the system.
A nonzero spectral gap often implies exponential decay of ground-state
correlation functions~\cite{Hastings2}. The
latter property implies the former if the system is Lorenz invariant,
but this is often not the case in condensed-matter physics.
Affleck, Kennedy, Lieb and Tasaki (AKLT) constructed a family of
two-dimensional spin models with isotropic spin-spin
interaction~\cite{AKLT_1988}, generalizing their spin-1 chain~\cite{AKLT_PRL},
now recognized as a paradigmatic example
of a symmetry protected topological phase. The ground state of the 2D AKLT
model on both the honeycomb and square lattices display exponentially
decaying correlations~\cite{AKLT_1988,KLT}.
However, the gappedness of the two models was not
proved rigorously for more than three decades.

Here, we close the loop by proving the existence of spectral gap in the
spin-3/2 AKLT models on four 2D Archimedean degree-3 lattices:
honeycomb, square-octagon, star and cross lattices.  For the original
spin-3/2 AKLT model, on the honeycomb lattice, the existence of a
spectral gap was independently established in a very recent work by
Lemm, Sandvik, and Wang~\cite{LemmSandvikWang} using a different, also
numerically-assisted, method. We also show the
gappedness of the hybrid AKLT models on the decorated honeycomb and the
decorated lattices, where a spin-1 site is added to every edge of the
undecorated lattice. These two cases were left open in recent studies of
AKLT models on decorated lattices~\cite{decorated1,decorated2}.
We note that the  AKLT models on the
above lattices except the honeycomb were not considered in the original
work, but they are as important as the honeycomb, as all of them provide
an example of  symmetry protected topological phases or Haldane phases,
as well as utility in realizing universal quantum computations using the
measurement-based approach
\cite{AKLT_QC_honeycomb,Miyake2011,Wei2013,Wei2014,AKLT_QC_square}.

Although the existence of the gap had not been proven, numerical tensor-network
methods were used to estimate its value~\cite{AKLTgap2013,Vanderstraeten2015,
Poilblanc2013},
which was shown to be of order $0.1$ for the honeycomb model. A combination of
analytics and numerics was recently used to show the spectral gap in the AKLT
hexagonal chain~\cite{hexagonalchain}. Consideration of decorated lattices has
recently led to analytic proofs of gaps for 2D AKLT models on the decorated
honeycomb lattice~\cite{decorated1}, and later on the decorated square and
other lattices~\cite{decorated2}. The latter two works left open the question
of whether AKLT models on the singly decorated honeycomb and square lattices
are also gapped, which we demonstrate postively below.

\section{Constructing Hamiltonian terms}
\label{sec:construction}
As in \cite{decorated1} and \cite{decorated2}, and based on Lemma~6.3 of
\cite{Fannes1992}, in order to establish the
existence of a gap we first extend the support of individual Hamiltonian terms.
To explain this in a uniform manner, we should introduce some notation.
For some subgraph $\Gamma$ of the full lattice $\Lambda$, 
\begin{itemize}
  \item We let $H_\Gamma$ be the AKLT Hamiltonian defined on that subgraph,
  \begin{equation}
    H_\Gamma = \sum_{e\in \Gamma}H^{(e)},
  \end{equation}
  where, for an edge $e$, $H^{(e)}$ projects onto the maximum spin subspace
  of the two vertices which $e$ joins, which corresponds to the combined
  coordination number $(z+z'+2)/2$.
  \item We let $\Psi_\Gamma$ be the AKLT construction on the subgraph
  $\Gamma$, obtained by taking the spin-$\frac{z}{2}$ projector on
  each vertex in $\Gamma$ (represented as a rank-$(z+1)$ tensor) and contracting
  with them the spin-singlet state on each edge in $\Gamma$.
  This leaves us with both virtual
  and physical indices; we group the former into a Hilbert space
  $\mathcal{H}_\text{virt}^\Gamma$ and the latter into a Hilbert space
  $\mathcal{H}_\text{phys}^\Gamma$, so that we can write $\Psi_\Gamma$ as 
  a linear transformation from $\mathcal{H}_\text{virt}^\Gamma$ to
  $\mathcal{H}_\text{phys}^\Gamma$.
  \item We write the singular value decomposition
  $\Psi_\Gamma = V_\Gamma\Sigma_\Gamma U_\Gamma^\dagger$, where we
  omit the trivial singular values, i.e. the space $\mathcal{H}_\Gamma$
  that $\Sigma_\Gamma$, $V_\Gamma$, and $U_\Gamma$ act on is truncated
  such that the latter is
  full-rank.
  \item With $\mathcal{H}_\Gamma$ defined as such,
  \begin{equation}
    \Pi_\Gamma \equiv U_\Gamma U_\Gamma^\dagger
  \end{equation}
  is the projector onto the image of $\Psi_\Gamma$, which ought to equal
  the kernel of $H_\Gamma$.
\end{itemize}
Then we will define new Hamiltonian terms, of the form
$\tilde{H}_i = \mathbbm{1}_{\Gamma_i,\text{phys}} - \Pi_{\Gamma_i}$,
by the subgraph $\Gamma_i$ on which the term is supported; in order to compare
this to the original AKLT Hamiltonian, we find a $\gamma_0$ such that
\begin{equation}
  \sum_{e \in \Gamma_i}\frac{1}{n_e}H^{(e)} \geq \gamma_0\tilde{H}_i,
  \label{eq:originalbound}
\end{equation}
where $n_e$ is the number of $i$ for which $e \in \Gamma_i$. (Note that
the frustration-freeness of the AKLT Hamiltonian allows us to vary the
coefficients of the terms comprising $H_\Gamma$ without altering the ground
space thereof.) In this case,
\begin{equation}
  H = \sum_{e \in \Lambda}H^{(e)} \geq \gamma_0 \sum_i\tilde{H}_i \equiv \tilde{H}
\end{equation}
bounds the original Hamiltonian $H$ with the new Hamiltonian $\tilde{H}$.

Having done this, we complete the bound by squaring the altered Hamiltonian:
if, for some $\tilde\gamma > 0$,
\begin{equation}
  \tilde{H}^2 \geq \tilde{\gamma}\tilde{H},
\end{equation}
then $\tilde{H}$ has a gap $\tilde{\gamma}$.

Since the $\tilde{H}_i$ are projectors, we can write
\begin{align}
    \tilde{H}^2 = \sum_{i,j}\tilde{H}_i\tilde{H}_j &= \sum_i\tilde{H}_i + \sum_{i\neq j}\tilde{H}_i\tilde{H}_j\\
    &\geq \tilde{H} + \sum_{\langle i,j\rangle}\{\tilde{H}_i,\tilde{H}_j\}
    \label{eq:nnsum}\\
    &\geq (1 - \eta z)\tilde{H},
    \label{eq:square2}
\end{align}
where the sum in \eqref{eq:nnsum} is over ``nearest neighbors''
$\langle i,j\rangle$ such that $\Gamma_i \cap \Gamma_j \neq \varnothing$, $z$
being the (maximal) number of nearest neighbors per term, and we define $\eta$
to be the maximal number such that
\begin{equation}
  \{E,F\} \equiv EF+FE \geq -\eta(E+F)
  \label{eq:anticommute}
\end{equation}
for all $E=\tilde{H}_i$ and $F=\tilde{H}_j$ nearest neighbors in the above
sense.

In \cite{decorated2}, we determine the following properties which we can
use to simplify \eqref{eq:anticommute}:
\begin{enumerate}
  \item The $\eta$ which is optimal in \eqref{eq:anticommute} for
  $E=\mathbbm{1}-\Pi_{\Gamma_i}$ and $F=\mathbbm{1}-\Pi_{\Gamma_j}$, then
  it is also optimal for $E=\Pi_{\Gamma_i}$ and $F=\Pi_{\Gamma_j}$.
  \item For $\eta$ optimal in \eqref{eq:anticommute}, we find that
  $1-\eta$ is the least noninteger eigenvalue of $E+F$; likewise
  $1+\eta$ is the greatest noninteger eigenvalue of $E+F$.
  \item If a projector $A=U_AU_A^\dagger$ commutes with both $E$ and $F$,
  in addition to which $EA = E$, then the ``reduced'' operator
  $U_A^\dagger(E+F)U_A$ has the same noninteger eigenvalues as $E+F$.
\end{enumerate}
The final element we need is the form of the projectors $A$. We first note
that, if $\Gamma\cap\Gamma'=\varnothing$, then, trivially,
$\Pi_\Gamma\otimes \mathbbm{1}_{\Gamma^c}$ commutes with
$\Pi_{\Gamma'}\otimes \mathbbm{1}_{\Gamma^{\prime c}}$. More subtly,

\begin{prop} If $\Gamma'\subset \Gamma$, then
$(\Pi_{\Gamma'}\otimes\mathbbm{1}_{\Gamma\setminus\Gamma'})\Pi_\Gamma=\Pi_\Gamma$.
\end{prop}

In addition to factorizing the physical space $\mathcal{H}_{\text{phys}}^\Gamma$
into $\mathcal{H}_0\equiv \mathcal{H}_{\text{phys}}^{\Gamma'}$ and
$\mathcal{H}_1\equiv \mathcal{H}_{\text{phys}}^{\Gamma\setminus\Gamma'}$,
we need to examine the interplay of virtual (bond) spaces. We can do so in a
straightforward way, by introducing an operator
$X:\mathcal{H}_{\text{virt}}^{\Gamma'}\otimes\mathcal{H}_{\text{virt}}^{\Gamma\setminus\Gamma'}\to\mathcal{H}_{\text{virt}}^\Gamma$ which corresponds
to contraction with the singlet on those pairs of indices which represent the
same edge, and which acts as the identity on all other indices. Then we can
write the AKLT construction on $\Gamma$ as
\begin{equation}
  \Psi_\Gamma = X(\Psi_{\Gamma'}\otimes\Psi_{\Gamma\setminus\Gamma'}).
\end{equation}
This implies, in particular, that the image of $\Psi_\Gamma$ (and therefore of
$\Pi_\Gamma$) is a subset of the image of
$\Psi_{\Gamma'}\otimes\mathbbm{1}_{\Gamma\setminus \Gamma'}$ (and therefore
of $\Pi_{\Gamma'}\otimes\mathbbm{1}_{\Gamma\setminus \Gamma'}$). The
relationship between these projectors follows immediately. $\square$

We then apply three simplifying projectors to $E$ and $F$, generated from
the disjoint subgraphs $L = \Gamma_i\setminus \Gamma_j$,
$R = \Gamma_j\setminus\Gamma_i$,
and $C=\Gamma_i\cap \Gamma_j$. That is, we use
\begin{align}
  U_E' &\equiv (U_L^\dagger\otimes U_C^\dagger)U_E:\mathcal{H}_L\otimes \mathcal{H}_C\to \mathcal{H}_E\notag\\
  U_F' &\equiv (U_R^\dagger\otimes U_C^\dagger)U_F:\mathcal{H}_R\otimes \mathcal{H}_C\to \mathcal{H}_F\\
  E+F &\mapsto U_E'U_E^{\prime\dagger}\otimes \mathbbm{1}_R + \mathbbm{1}_L\otimes U_F'U_F^{\prime\dagger}
\end{align}

For purposes below, we will also consider the set $Y=\Gamma_i\cup\Gamma_j$,
which we mention here to note that the edge set of $Y$ is not guaranteed to
contain all edges joining the vertices of $Y$, in contrast to the other
subgraphs described above.

We further note that we can easily determine the dimension of this new space
$\mathcal{H}_\Gamma$ corresponding to the singular values of some $\Psi_\Gamma$.
Looking at the collection of virtual indices comprising
$\mathcal{H}_\text{virt}^\Gamma$, we group together those originating from the
same site and symmetrize them, which reduces their collective dimension from
$2^k$ to $k+1$. Then the domain of the resulting space can be obtained as
follows: for each vertex $v\in\Gamma$, let $k_v$ be the number of ``free''
indices, that is, edges of $\Lambda$ which terminate in $v$ but which do
not belong to the subgraph $\Gamma$. Then we expect
\begin{equation}
  \dim\mathcal{H}_\Gamma = \prod_{v\in\Gamma}(k_v+1).
\end{equation}
We can confirm equality
on a case-by-case basis when performing the singular value decomposition, and,
when doing so, the upper bound allows us to rule out the possibility 
that positive singular values have been improperly discarded due to being
below machine precision.

Having completed these preliminaries, we will lay out the three different ways
we use these tensors to extract the number $\eta$:
\begin{enumerate}[I]
  \item \textit{Exact diagonalization}: We construct $E'$ and $F'$
  explicitly and diagonalize $E'+F'$, whose eigenvalues are in
  $[0,2]$. As apparently-integer eigenvalues will
  only be given up to machine precision, we confirm the apparent eigenspace
  of 2 by noting that the 2-eigenvectors of $E+F$ should correspond to
  the image of $\Psi_Y$, and compare the apparent degeneracy of 2 with
  the expected dimension of this image. Then we can obtain the greatest
  non-integer eigenvalue of $E'+F'$, which will equal $1+\eta$.
  We use this method to demonstrate the existence of the
  gap on the square-octagon, star, and cross lattices.
  \item \textit{Iterative diagonalization with} $\Pi_Y'$: When 
  $\mathcal{H}_L\otimes\mathcal{H}_C\otimes\mathcal{H}_R$ is too large for us
  to construct, and therefore diagonalize, $E'+F'$, we instead iteratively
  diagonalize it after ``shifting'' the eigenspace of 2: We note, again,
  that the 2-eigenspace of $E+F$ should equal the image of $\Psi_Y$,
  or equivalently, that of $E'+F'$ will equal the image of
  $\Psi_Y(U_L\otimes U_C\otimes U_R)$. In particular, the projector
  $\Pi_Y'$ will project onto the 2-eigenspace of $E'+F'$; therefore,
  as long as $E'+F'$ has any eigenvalues in $(1,2)$,
  the greatest eigenvalue of $E'+F'-\Pi_Y'$ will be the greatest
  \textit{noninteger} eigenvalue of $E'+F'$, i.e. $1+\eta$. Therefore, we can
  use Lanczos diagonalization procedures (in particular, the high-precision
  implementation provided by ARPACK) which select the greatest-magnitude
  eigenvectors of a Hermitian operator. We use this method to demonstrate the
  existence of the gap on the singly-decorated square and honeycomb lattices.
  \item \textit{Iterative diagonalization with} $\Psi_Y'$: When we cannot
  explicitly construct $\Psi_Y'$, and therefore cannot obtain $U_Y'$,
  we can still apply $\Psi_Y'$ to vectors in
  $\mathcal{H}_L\otimes\mathcal{H}_C\otimes\mathcal{H}_R$: in particular
  we can construct an operator $\rho_Y\equiv \Psi_Y^{\prime\dagger}\Psi_Y'$
  in order to ``shift'' the image of $\Psi_Y'$ as above. In particular, we
  may seek the greatest eigenvalues of the following operators:
  \begin{align}
      O_1 &\equiv -(E'F'+F'E')\notag\\
      O_2 &\equiv \frac{5}{2}(E'+F') - (E'F'+F'E') - \varepsilon\rho_Y
      \label{eq:iterative-ops}
  \end{align}
  where $\varepsilon$ can be tuned as needed (we will choose a value of 0.1).
  Identifying the eigenvalues of $E'+F'$, as usual, by $1-\alpha$,
  the eigenvalues of $O_1$ are $\alpha(1-\alpha)$, which is maximized by
  $\alpha=\frac{1}{2}$: in particular, if $\eta\leq\frac{1}{2}$, then the
  greatest eigenvalue of $O_1$ will be $\eta(1-\eta)$. Meanwhile,
  for $\alpha \neq -1$, the eigenvalues of $O_2$ are
  $(\alpha+\frac{5}{2})(1-\alpha)$, which is maximized by
  $\alpha=-\frac{3}{4}$. In particular, if there are any
  $\alpha \in (-1,-\frac{1}{2})$, the greatest eigenvalue of $O_2$ will
  exceed 3; meanwhile, if $\eta < \frac{1}{2}$, we ensure by subtracting
  a multiple of $\rho_Y$ that the greatest eigenvalue of $O_2$ is
  \textit{strictly less than} 3. We use this method to demonstrate the existence
  of the AKLT gap on the honeycomb lattice.
\end{enumerate}

\section{Re-partitioning lattices}
A summary of values of the key parameter $\eta$, relevant dimensions, and
other information is shown in Table~\ref{tab:data}.
\subsection{The honeycomb lattice}
\begin{figure} 
  \begin{subfigure}
    \centering
    \begin{tikzpicture}[scale=1.00, every node/.style={transform shape}]
      \foreach \i in {0,...,8}
      \foreach \j in {0,...,11} {
        \node[coordinate] (\i\j a) at (3/2*\i,3*sin{60}*\j) {};
        \node[coordinate] (\i\j b) at (3/2*\i+.75,3*sin{60}*\j+3/2*sin{60}) {};
        \foreach \a in {0,60,120,180,240,300}
        \foreach \k in {a,b} {
          \node[coordinate] (\i\j\k;\a) at ($(\i\j\k)+(\a:1/2)$) {};
        }
      }
      \clip(11b) rectangle (74b);
      \foreach \i in {0,...,7}
      \foreach \j in {0,...,10} {
        \foreach \k in {a,b} {
          \foreach \a in {0,60,120,180,240,300} {
            \node[circle,fill=black,inner sep=0pt,minimum size=1mm] (\i\j\k\a;p) at (\i\j\k;\a) {}; }
          \draw (\i\j\k;0) -- (\i\j\k;60) -- (\i\j\k;120) -- (\i\j\k;180) -- (\i\j\k;240) -- (\i\j\k;300) -- (\i\j\k;0);
        }
        \draw (\i\j a;60) -- (\i\j b;240);
        \pgfmathtruncatemacro\n{\i+1}
        \pgfmathtruncatemacro\m{\j+1}
        \draw (\i\j a;0) -- (\n\j a;180);
        \draw (\i\j b;300) -- (\n\j a;120);
        \draw (\i\j b;0) -- (\n\j b;180);
        \draw (\i\j b;120) -- (\i\m a;300);
        \draw (\i\j b;60) -- (\n\m a;240);
      }
      \foreach \i in {1,...,3}
      \foreach \j in {1,...,9} {
        \pgfmathtruncatemacro\ia{2*\i-1}
        \pgfmathtruncatemacro\ib{2*\i}
        \pgfmathtruncatemacro\ic{2*\i+1}
        \pgfmathtruncatemacro\id{2*\i+2}
        \pgfmathtruncatemacro\ja{\j-1}
        \pgfmathtruncatemacro\jb{\j}
        \pgfmathtruncatemacro\jc{\j+1}
        \draw[cyan,dashed,very thick] plot [smooth cycle, tension=.7] coordinates {
          ($(\ib\jb a)+(210:.8)$)
          ($(\ia\jb a;0)  +(240:.14)$)  ($(\ia\jb a;60) +(180:.08)$)
          ($(\ia\jb b;240)+(120:.08)$)  ($(\ia\jb b;300)+ (60:.14)$)
          ($(\ib\jb a)+(90:.8)$)
          ($(\ib\jb b;240)+(120:.14)$)  ($(\ib\jb b;300)+ (60:.08)$)
          ($(\ic\jb a;120)+  (0:.08)$)  ($(\ic\jb a;180)+(300:.14)$)
          ($(\ib\jb a)+(330:.8)$)
          ($(\ib\ja b;120)+  (0:.14)$)  ($(\ib\ja b;180)+(300:.08)$)
          ($(\ia\ja b;0)  +(240:.08)$)  ($(\ia\ja b;60) +(180:.14)$) };
        \draw[red,dashed,very thick] plot [smooth cycle, tension=.7] coordinates {
          ($(\ic\jb a)+(210:.8)$)
          ($(\ib\jb a;0)  +(240:.14)$)  ($(\ib\jb a;60) +(180:.08)$)
          ($(\ib\jb b;240)+(120:.08)$)  ($(\ib\jb b;300)+ (60:.14)$)
          ($(\ic\jb a)+(90:.8)$)
          ($(\ic\jb b;240)+(120:.14)$)  ($(\ic\jb b;300)+ (60:.08)$)
          ($(\id\jb a;120)+  (0:.08)$)  ($(\id\jb a;180)+(300:.14)$)
          ($(\ic\jb a)+(330:.8)$)
          ($(\ic\ja b;120)+  (0:.14)$)  ($(\ic\ja b;180)+(300:.08)$)
          ($(\ib\ja b;0)  +(240:.08)$)  ($(\ib\ja b;60) +(180:.14)$) };
        \draw[violet,dashed,very thick] plot [smooth cycle, tension=.7] coordinates {
          ($(\ib\jb b)+(210:.8)$)
          ($(\ia\jb b;0)  +(240:.14)$)  ($(\ia\jb b;60) +(180:.08)$)
          ($(\ib\jc a;240)+(120:.08)$)  ($(\ib\jc a;300)+ (60:.14)$)
          ($(\ib\jb b)+(90:.8)$)
          ($(\ic\jc a;240)+(120:.14)$)  ($(\ic\jc a;300)+ (60:.08)$)
          ($(\ic\jb b;120)+  (0:.08)$)  ($(\ic\jb b;180)+(300:.14)$)
          ($(\ib\jb b)+(330:.8)$)
          ($(\ic\jb a;120)+  (0:.14)$)  ($(\ic\jb a;180)+(300:.08)$)
          ($(\ib\jb a;0)  +(240:.08)$)  ($(\ib\jb a;60) +(180:.14)$) };
        \draw[green,dashed,very thick] plot [smooth cycle, tension=.7] coordinates {
          ($(\ic\jb b)+(210:.8)$)
          ($(\ib\jb b;0)  +(240:.14)$)  ($(\ib\jb b;60) +(180:.08)$)
          ($(\ic\jc a;240)+(120:.08)$)  ($(\ic\jc a;300)+ (60:.14)$)
          ($(\ic\jb b)+(90:.8)$)
          ($(\id\jc a;240)+(120:.14)$)  ($(\id\jc a;300)+ (60:.08)$)
          ($(\id\jb b;120)+  (0:.08)$)  ($(\id\jb b;180)+(300:.14)$)
          ($(\ic\jb b)+(330:.8)$)
          ($(\id\jb a;120)+  (0:.14)$)  ($(\id\jb a;180)+(300:.08)$)
          ($(\ic\jb a;0)  +(240:.08)$)  ($(\ic\jb a;60) +(180:.14)$) };
      }
    \node[below] at (current bounding box.south) {\fontsize{8}{8}\selectfont (b)};
    \end{tikzpicture}
  \end{subfigure}
  \begin{subfigure}
    \centering
    \begin{tikzpicture}[scale=1.50]
      \foreach \i in {0,...,8}
      \foreach \j in {0,...,11} {
        \node[coordinate] (\i\j a) at (3/2*\i,3*sin{60}*\j) {};
        \node[coordinate] (\i\j b) at (3/2*\i+.75,3*sin{60}*\j+3/2*sin{60}) {};
        \foreach \a in {0,60,120,180,240,300}
        \foreach \k in {a,b} {
          \node[coordinate] (\i\j\k;\a) at ($(\i\j\k)+(\a:1/2)$) {};
        }
      }
      \node[Lv] (Lt1) at (11a;180) {};
      \node[Lv] (Lt2) at (01a;0) {};
      \node[Lv] (Lt3) at (01a;60) {};
      \node[Lv] (Lt4) at (01b;240) {};
      \node[Lv] (Lt5) at (01b;300) {};
      \node[Lv] (Lt6) at (11a;120) {};
      \node[Lv] (Lb1) at (11a;240) {};
      \node[Lv] (Lb2) at (00b;60) {};
      \node[Lv] (Lb3) at (00b;0) {};
      \node[Lv] (Lb4) at (10b;180) {};
      \node[Lv] (Lb5) at (10b;120) {};
      \node[Lv] (Lb6) at (11a;300) {};
      \draw[blue] (Lt1) -- (Lt2) -- (Lt3) -- (Lt4) -- (Lt5) -- (Lt6) -- (Lt1) -- (Lb1) -- (Lb2) -- (Lb3) -- (Lb4) -- (Lb5) -- (Lb6) -- (Lb1);
      \node[Cv] (C1) at (21a;120) {};
      \node[Cv] (C2) at (11b;300) {};
      \node[Cv] (C3) at (11b;240) {};
      \node[Cv] (C4) at (11a;60) {};
      \node[Cv] (C5) at (11a;0) {};
      \node[Cv] (C6) at (21a;180) {};
      \draw[red] (C1) -- (C2) -- (C3) -- (C4) -- (C5) -- (C6) -- (C1);
      \draw[blue,dashed] (C4) -- (Lt6);
      \draw[blue,dashed] (C5) -- (Lb6);
      \node[Rv] (Rt1) at (21a;0) {};
      \node[Rv] (Rt2) at (31a;180) {};
      \node[Rv] (Rt3) at (31a;120) {};
      \node[Rv] (Rt4) at (21b;300) {};
      \node[Rv] (Rt5) at (21b;240) {};
      \node[Rv] (Rt6) at (21a;60) {};
      \node[Rv] (Rb1) at (21a;300) {};
      \node[Rv] (Rb2) at (20b;120) {};
      \node[Rv] (Rb3) at (20b;180) {};
      \node[Rv] (Rb4) at (10b;0) {};
      \node[Rv] (Rb5) at (10b;60) {};
      \node[Rv] (Rb6) at (21a;240) {};
      \draw[green] (Rt1) -- (Rt2) -- (Rt3) -- (Rt4) -- (Rt5) -- (Rt6) -- (Rt1) -- (Rb1) -- (Rb2) -- (Rb3) -- (Rb4) -- (Rb5) -- (Rb6) -- (Rb1);
      \draw[green,dashed] (C1) -- (Rt6);
      \draw[green,dashed] (C6) -- (Rb6);
    \end{tikzpicture}
  \end{subfigure}
  \caption{Above: Enlarged terms of the honeycomb lattice, described below.
  Below: pairs of adjacent terms, overlapping on hexagonal plaquettes.
  Blue, green, and red represent the regions on which we apply the
  ``simplifying'' isometries $U_L$, $U_R$, and $U_C$, respectively.}
  \label{fig:honeycomb}
\end{figure}
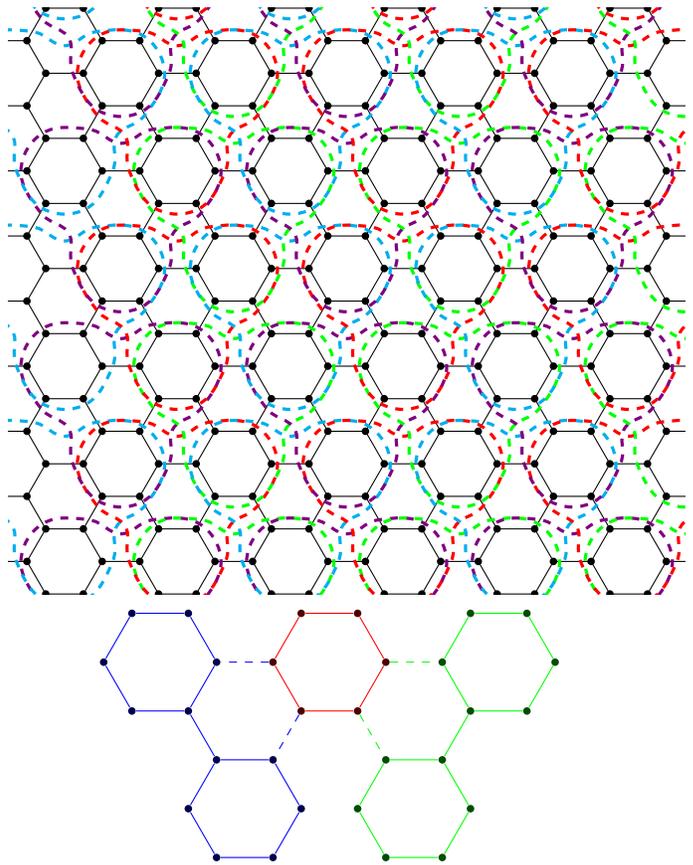
To find plaquettes we can use to demonstrate the existence of the gap on the
honeycomb lattice, we first tripartition the dual lattice; call the resultant
sets of plaquettes $\mathcal{A}$, $\mathcal{B}$, and $\mathcal{C}$. Then,
as shown in Fig.~\ref{fig:honeycomb}, we assign for each plaquette
$p\in \mathcal{A}$ a subgraph $\Gamma_p$ consisting of $p$ and the three
neighboring plaquettes belonging to $\mathcal{B}$. Then the overlapping
subgraphs consist of nearest neighbors of the triangular lattice whose vertices
are the elements of $\mathcal{A}$; in particular, each term overlaps with
6 other terms, so we must find $\eta < \frac{1}{6}$.
As shown, the overlapping subgraph $C$ will
be a hexagon, whereas the outside subgraphs $L$ and $R$ consist of
nonintersecting hexagonal plaquettes joined by a single edge. These allow
us to reduce the dimension of the space which $E'+F'$ acts on to
\begin{align}
\dim\mathcal{H}_C &= 2^6\notag\\
\dim\mathcal{H}_L = \dim\mathcal{H}_R &= 2^{10}\notag\\
\dim(\mathcal{H}_L\otimes\mathcal{H}_C\otimes\mathcal{H}_R) &= 2^{26}
\end{align}
Then we determine that
$\eta = 0.1445124916 < \frac{1}{6}$.
\subsection{The [4.8.8] square-octagon lattice}
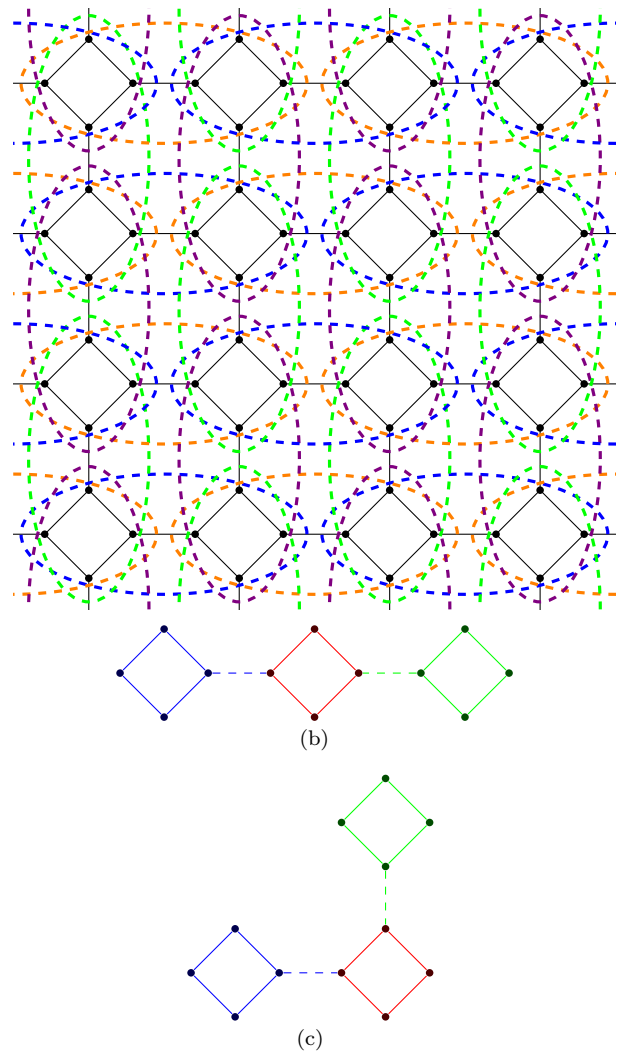
\begin{figure} 
  \centering
  \begin{subfigure}
    \centering
    \begin{tikzpicture}[scale=1.00, every node/.style={transform shape}]
      \foreach \i in {0,...,8}
      \foreach \j in {0,...,8} {
        \node[coordinate] (\i\j) at (2*\i,2*\j) {};
        \foreach \a in {0,90,180,270} {
          \node[coordinate] (\i\j;\a) at ($(\i\j)+(\a:2-2*sin{45})$) {};
        }
      }
      \clip($(00)!.5!(11)$) rectangle ($(44)!.5!(55)$);
      \foreach \i in {0,...,7}
      \foreach \j in {0,...,7} {
        \foreach \a in {0,90,180,270} {
          \node[circle,fill=black,inner sep=0pt,minimum size=1mm] (\i\j;\a p) at (\i\j;\a) {}; }
        \draw (\i\j;0) -- (\i\j;90) -- (\i\j;180) -- (\i\j;270) -- (\i\j;0);
        \pgfmathtruncatemacro\n{\i+1}
        \pgfmathtruncatemacro\m{\j+1}
        \draw (\i\j;0) -- (\n\j;180);
        \draw (\i\j;90) -- (\i\m;270);
        }
      \foreach \i in {0,...,3}
      \foreach \j in {0,...,3} {
        \pgfmathtruncatemacro\ia{2*\i}
        \pgfmathtruncatemacro\ib{2*\i+1}
        \pgfmathtruncatemacro\ic{2*\i+2}
        \pgfmathtruncatemacro\ja{2*\j}
        \pgfmathtruncatemacro\jb{2*\j+1}
        \pgfmathtruncatemacro\jc{2*\j+2}
        \draw[dashed,very thick,draw=blue] ($(\ia\ja)!.5!(\ib\ja)$) ellipse (1.9 and .8);
        \draw[dashed,very thick,draw=blue] ($(\ib\jb)!.5!(\ic\jb)$) ellipse (1.9 and .8);
        \draw[dashed,very thick,draw=orange] ($(\ib\ja)!.5!(\ic\ja)$) ellipse (1.9 and .8);
        \draw[dashed,very thick,draw=orange] ($(\ia\jb)!.5!(\ib\jb)$) ellipse (1.9 and .8);
        \draw[dashed,very thick,draw=green] ($(\ia\ja)!.5!(\ia\jb)$) ellipse (.8 and 1.9);
        \draw[dashed,very thick,draw=green] ($(\ib\jb)!.5!(\ib\jc)$) ellipse (.8 and 1.9);
        \draw[dashed,very thick,draw=violet] ($(\ib\ja)!.5!(\ib\jb)$) ellipse (.8 and 1.9);
        \draw[dashed,very thick,draw=violet] ($(\ia\jb)!.5!(\ia\jc)$) ellipse (.8 and 1.9);
      }
    \node[below] at (current bounding box.south) {\fontsize{8}{8}\selectfont (a)};
    \end{tikzpicture}
  \end{subfigure}
  \begin{subfigure}
    \centering
    \begin{tikzpicture}[scale=1.00, every node/.style={transform shape}]
      \foreach \i in {0,...,8}
      \foreach \j in {0,...,8} {
        \node[coordinate] (\i\j) at (2*\i,2*\j) {};
        \foreach \a in {0,90,180,270} {
          \node[coordinate] (\i\j;\a) at ($(\i\j)+(\a:2-2*sin{45})$) {};
        }
      }
      \node[Lv] (L1) at (00;0) {};
      \node[Lv] (L2) at (00;90) {};
      \node[Lv] (L3) at (00;180) {};
      \node[Lv] (L4) at (00;270) {};
      \node[Cv] (C1) at (10;0) {};
      \node[Cv] (C2) at (10;90) {};
      \node[Cv] (C3) at (10;180) {};
      \node[Cv] (C4) at (10;270) {};
      \node[Rv] (R1) at (20;0) {};
      \node[Rv] (R2) at (20;90) {};
      \node[Rv] (R3) at (20;180) {};
      \node[Rv] (R4) at (20;270) {};
      \draw[blue] (L1) -- (L2) -- (L3) -- (L4) -- (L1);
      \draw[blue,dashed] (L1) -- (C3);
      \draw[red] (C1) -- (C2) -- (C3) -- (C4) -- (C1);
      \draw[green] (R1) -- (R2) -- (R3) -- (R4) -- (R1);
      \draw[green,dashed] (R3) -- (C1);
    \node[below] at (current bounding box.south) {\fontsize{8}{8}\selectfont (b)};
    \end{tikzpicture}
  \end{subfigure}
  \begin{subfigure}
    \centering
    \begin{tikzpicture}[scale=1.00, every node/.style={transform shape}]
      \foreach \i in {0,...,8}
      \foreach \j in {0,...,8} {
        \node[coordinate] (\i\j) at (2*\i,2*\j) {};
        \foreach \a in {0,90,180,270} {
          \node[coordinate] (\i\j;\a) at ($(\i\j)+(\a:2-2*sin{45})$) {};
        }
      }
      \node[Lv] (L1) at (00;0) {};
      \node[Lv] (L2) at (00;90) {};
      \node[Lv] (L3) at (00;180) {};
      \node[Lv] (L4) at (00;270) {};
      \node[Cv] (C1) at (10;0) {};
      \node[Cv] (C2) at (10;90) {};
      \node[Cv] (C3) at (10;180) {};
      \node[Cv] (C4) at (10;270) {};
      \node[Rv] (R1) at (11;0) {};
      \node[Rv] (R2) at (11;90) {};
      \node[Rv] (R3) at (11;180) {};
      \node[Rv] (R4) at (11;270) {};
      \draw[blue] (L1) -- (L2) -- (L3) -- (L4) -- (L1);
      \draw[blue,dashed] (L1) -- (C3);
      \draw[red] (C1) -- (C2) -- (C3) -- (C4) -- (C1);
      \draw[green] (R1) -- (R2) -- (R3) -- (R4) -- (R1);
      \draw[green,dashed] (R4) -- (C2);
    \node[below] at (current bounding box.south) {\fontsize{8}{8}\selectfont (c)};
    \end{tikzpicture}
  \end{subfigure}
  \caption{Above: Enlarged terms of the square-octagon lattice, corresponding
  to neighboring pairs of square plaquettes. Below: overlapping terms for
  which the two octagon-octagon edges are (b) collinear or (c) perpendicular.
  }
  \label{fig:squoct}
\end{figure}
The new terms we consider correspond to an edge dividing a given pair of
octogonal plaquettes and the two square plaquettes it connects. Here each
such subgraph overlaps with six others; uniquely among the lattices we are
considering, the overlaps between these subgraphs have two different forms.
We categorize these terms ($\Gamma_i$ and $\Gamma_j$)
using the edges connecting the square plaquettes, in that
they can be either collinear (Fig.~\ref{fig:squoct}b, say $i \parallel j$) or
perpendicular (Fig.~\ref{fig:squoct}c, say $i \perp j$). We must then
modify \eqref{eq:anticommute} and \eqref{eq:square2} to read
\begin{align}
  \{\tilde{H}_i,\tilde{H}_j\} &\geq -\eta_\parallel(\tilde{H}_i+\tilde{H}_j),\ i\parallel j\notag\\
    &\geq -\eta_\perp(\tilde{H}_i+\tilde{H}_j),\ i\perp j\\
  \tilde{H}^2 &\geq (1-2\eta_\parallel-4\eta_\perp)\tilde{H}.
\end{align}
In particular, we need to demonstrate $4\eta_\perp~+~2\eta_\parallel~<~1$.
We determine that $\eta_\parallel=0.1061446858$ and
$\eta_\perp=0.1589663310$.

\subsection{The [3.12.12] ``star'' lattice}
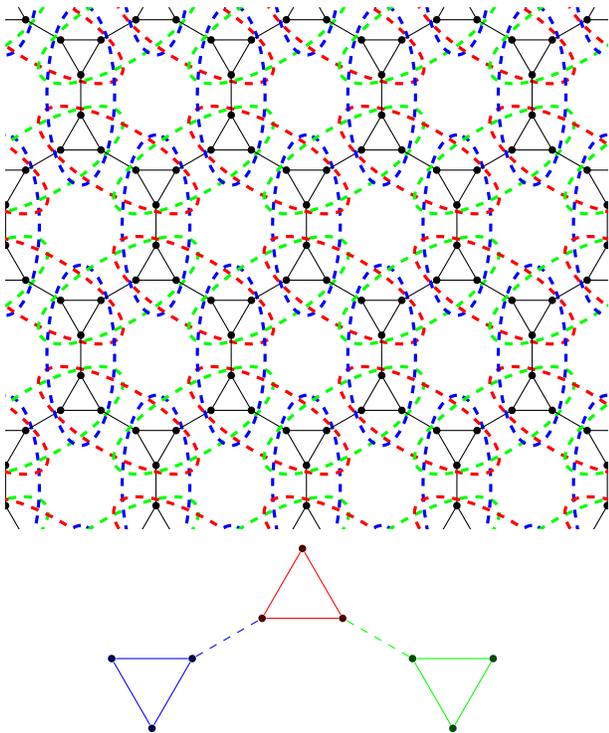
\begin{figure} 
  \begin{subfigure}
    \centering
    \begin{tikzpicture}[scale=1.00, every node/.style={transform shape}]
      \foreach \i in {0,...,7}
      \foreach \j in {0,...,3} {
        \node[coordinate] (\i\j a0) at (2*\i,4*sin{60}*\j) {};
        \node[coordinate] (\i\j a1) at (2*\i,4*sin{60}*\j+4/3*sin{60}) {};
        \node[coordinate] (\i\j b0) at (2*\i+2*cos{60},4*sin{60}*\j+2*sin{60}) {};
        \node[coordinate] (\i\j b1) at (2*\i+2*cos{60},4*sin{60}*\j+10/3*sin{60}) {};
        \foreach \a in {90,210,330}
        \foreach \k in {a,b} {
          \node[coordinate] (\i\j\k 0\a) at ($(\i\j\k 0)+(\a:8/3*sin{60}-2)$) {};
          \node[coordinate] (\i\j\k 1\a) at ($(\i\j\k 1)+(\a:-8/3*sin{60}+2)$) {}; }
        }
      \clip(00b0) rectangle (42b0);
      \foreach \i in {0,...,6}
      \foreach \j in {0,...,2} {
        \foreach \k in {a,b}
        \foreach \l in {0,1} {
          \foreach \a in {90,210,330}
            \node[circle,fill=black,inner sep=0pt,minimum size=1mm] (\i\j\k\l\a p) at (\i\j\k\l\a) {};
          \draw (\i\j\k\l 90) -- (\i\j\k\l 210) -- (\i\j\k\l 330) -- (\i\j\k\l 90); }
        \draw (\i\j a090) -- (\i\j a190);
        \draw[rotate=90,dashed,very thick,draw=blue] ($(\i\j a0)!.5!(\i\j a1)$) ellipse (1.2 and .45);
        \draw (\i\j a1210) -- (\i\j b0210);
        \draw[rotate=210,dashed,very thick,draw=green] ($(\i\j a1)!.5!(\i\j b0)$) ellipse (1.2 and .45);
        \draw (\i\j b090) -- (\i\j b190);
        \draw[rotate=90,dashed,very thick,draw=blue] ($(\i\j b0)!.5!(\i\j b1)$) ellipse (1.2 and .45);
        \pgfmathtruncatemacro\n{\i+1}
        \pgfmathtruncatemacro\m{\j+1}
        \draw (\i\j b0330) -- (\n\j a1330);
        \draw[rotate=330,dashed,very thick,draw=red] ($(\i\j a0)!.5!(\n\j b1)$) ellipse (1.2 and .45);
        \draw (\i\m a0330) -- (\i\j b1330);
        \draw[rotate=330,dashed,very thick,draw=red] ($(\i\m a0)!.5!(\i\j b1)$) ellipse (1.2 and .45);
        \draw (\i\j b1210) -- (\n\m a0210);
        \draw[rotate=210,dashed,very thick,draw=green] ($(\i\j b1)!.5!(\n\m a0)$) ellipse (1.2 and .45);
      }
    \node[below] at (current bounding box.south) {\fontsize{8}{8}\selectfont (b)};
    \end{tikzpicture}
  \end{subfigure}
  \begin{subfigure}
    \centering
    \begin{tikzpicture}[scale=2.00]
      \foreach \i in {0,...,7}
      \foreach \j in {0,...,3} {
        \node[coordinate] (\i\j a0) at (2*\i,4*sin{60}*\j) {};
        \node[coordinate] (\i\j a1) at (2*\i,4*sin{60}*\j+4/3*sin{60}) {};
        \node[coordinate] (\i\j b0) at (2*\i+2*cos{60},4*sin{60}*\j+2*sin{60}) {};
        \node[coordinate] (\i\j b1) at (2*\i+2*cos{60},4*sin{60}*\j+10/3*sin{60}) {};
        \foreach \a in {90,210,330}
        \foreach \k in {a,b} {
          \node[coordinate] (\i\j\k 0\a) at ($(\i\j\k 0)+(\a:8/3*sin{60}-2)$) {};
          \node[coordinate] (\i\j\k 1\a) at ($(\i\j\k 1)+(\a:-8/3*sin{60}+2)$) {}; }
        }
      \node[Lv] (L1) at (00a190) {};
      \node[Lv] (L2) at (00a1210) {};
      \node[Lv] (L3) at (00a1330) {};
      \draw[blue] (L1) -- (L2) -- (L3) -- (L1);
      \node[Cv] (C1) at (00b090) {};
      \node[Cv] (C2) at (00b0210) {};
      \node[Cv] (C3) at (00b0330) {};
      \draw[red] (C1) -- (C2) -- (C3) -- (C1);
      \node[Rv] (R1) at (10a190) {};
      \node[Rv] (R2) at (10a1210) {};
      \node[Rv] (R3) at (10a1330) {};
      \draw[green] (R1) -- (R2) -- (R3) -- (R1);
      \draw[blue,dashed] (L2) -- (C2);
      \draw[green,dashed] (R3) -- (C3);
    \end{tikzpicture}
  \end{subfigure}
  \caption{Above: Enlarged terms of the star lattice, corresponding
  to neighboring pairs of triangular plaquettes. Below: pairs of
  adjacent terms, overlapping on triangular plaquettes.}
  \label{fig:star}
\end{figure}
Similarly to the square-octagon lattice, in considering the ``star'' lattice we
select subgraphs consisting of the edge dividing a pair of dodecagonal
plaquettes and the two triangular plaquettes it connects. Now each such
subgraph overlaps with four others, and we find that
$\eta=0.1110430220<\frac{1}{4}$.

\subsection{The [4.6.12] ``cross'' lattice}
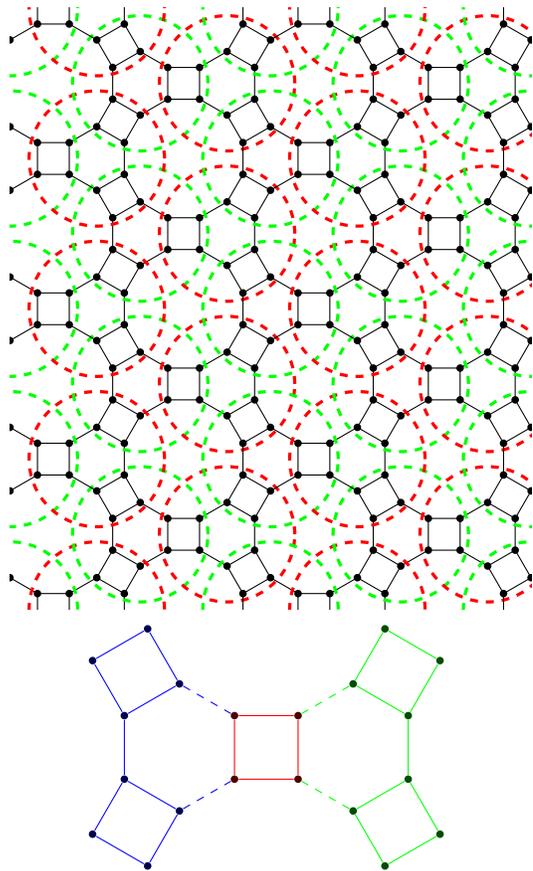
\begin{figure} 
  \begin{subfigure}
    \centering
    \begin{tikzpicture}[scale=1.00, every node/.style={transform shape}]
      \foreach \i in {0,...,5}
      \foreach \j in {0,...,7} {
        \node[coordinate] (\i\j a0) at (4*sin{60}*\i,2*\j) {};
        \node[coordinate] (\i\j a1) at (4*sin{60}*\i+4/3*sin{60},2*\j) {};
        \node[coordinate] (\i\j b0) at (4*sin{60}*\i+2*sin{60},2*\j+2*cos{60}) {};
        \node[coordinate] (\i\j b1) at (4*sin{60}*\i+10/3*sin{60},2*\j+2*cos{60}) {};
        \foreach \a in {30,90,150,210,270,330}
        \foreach \k in {a,b}
        \foreach \l in {0,1} {
          \node[coordinate] (\i\j\k\l\a) at ($(\i\j\k\l)+(\a:1-2/3*sin{60})$) {}; }
        }
      \clip(00a0) rectangle (24a0);
      \foreach \i in {0,...,4}
      \foreach \j in {0,...,6} {
        \foreach \k in {a,b}
        \foreach \l in {0,1} {
          \foreach \a in {30,90,150,210,270,330} {
            \node[circle,fill=black,inner sep=0pt,minimum size=1mm] (\i\j\k\l\a p) at (\i\j\k\l\a) {}; }
          \draw (\i\j\k\l 30) -- (\i\j\k\l 90) -- (\i\j\k\l 150) -- (\i\j\k\l 210) -- (\i\j\k\l 270) -- (\i\j\k\l 330) -- (\i\j\k\l 30);
        }
        \foreach \k in {a,b} {
          \draw[dashed,very thick,draw=green] (\i\j\k 0) circle (.9);
          \draw[dashed,very thick,draw=red] (\i\j\k 1) circle (.9);
        }
        \draw (\i\j a030) -- (\i\j a1150);
        \draw (\i\j a0330) -- (\i\j a1210);
        \draw (\i\j a190) -- (\i\j b0210);
        \draw (\i\j a130) -- (\i\j b0270);
        \draw (\i\j b030) -- (\i\j b1150);
        \draw (\i\j b0330) -- (\i\j b1210);
        \pgfmathtruncatemacro\n{\i+1}
        \pgfmathtruncatemacro\m{\j+1}
        \draw (\i\j b0150) -- (\i\m a1270);
        \draw (\i\j b090) -- (\i\m a1330);
        \draw (\n\j a0150) -- (\i\j b1270);
        \draw (\n\j a090) -- (\i\j b1330);
        \draw (\i\j b190) -- (\n\m a0210);
        \draw (\i\j b130) -- (\n\m a0270);
      }
    \node[below] at (current bounding box.south) {\fontsize{8}{8}\selectfont (c)};
    \end{tikzpicture}
  \end{subfigure}
  \begin{subfigure}
    \centering
    \begin{tikzpicture}[scale=2.00]
      \foreach \i in {0,...,3}
      \foreach \j in {0,...,7} {
        \node[coordinate] (\i\j a0) at (4*sin{60}*\i,2*\j) {};
        \node[coordinate] (\i\j a1) at (4*sin{60}*\i+4/3*sin{60},2*\j) {};
        \node[coordinate] (\i\j b0) at (4*sin{60}*\i+2*sin{60},2*\j+2*cos{60}) {};
        \node[coordinate] (\i\j b1) at (4*sin{60}*\i+10/3*sin{60},2*\j+2*cos{60}) {};
        \foreach \a in {30,90,150,210,270,330}
        \foreach \k in {a,b}
        \foreach \l in {0,1} {
          \node[coordinate] (\i\j\k\l\a) at ($(\i\j\k\l)+(\a:1-2/3*sin{60})$) {}; }
        }
      \node[Lv] (Lb2) at (00a190) {};
      \node[Lv] (Lb3) at (00a130) {};
      \node[Lv] (Lb4) at (00b0270) {};
      \node[Lv] (Lb1) at (00b0210) {};
      \node[Lv] (Lt2) at (01a1270) {};
      \node[Lv] (Lt3) at (01a1330) {};
      \node[Lv] (Lt4) at (00b090) {};
      \node[Lv] (Lt1) at (00b0150) {};
      \draw[blue] (Lb1) -- (Lb2) -- (Lb3) -- (Lb4) -- (Lb1) -- (Lt1) -- (Lt2) -- (Lt3) -- (Lt4) -- (Lt1);
      \node[Cv] (Cbl) at (00b0330) {};
      \node[Cv] (Ctl) at (00b030) {};
      \node[Cv] (Ctr) at (00b1150) {};
      \node[Cv] (Cbr) at (00b1210) {};
      \draw[blue,dashed] (Lb4) -- (Cbl);
      \draw[blue,dashed] (Lt4) -- (Ctl);
      \draw[red] (Cbl) -- (Cbr) -- (Ctr) -- (Ctl) -- (Cbl);
      \node[Rv] (Rb2) at (10a090) {};
      \node[Rv] (Rb3) at (10a0150) {};
      \node[Rv] (Rb4) at (00b1270) {};
      \node[Rv] (Rb1) at (00b1330) {};
      \node[Rv] (Rt2) at (11a0270) {};
      \node[Rv] (Rt3) at (11a0210) {};
      \node[Rv] (Rt4) at (00b190) {};
      \node[Rv] (Rt1) at (00b130) {};
      \draw[green] (Rb1) -- (Rb2) -- (Rb3) -- (Rb4) -- (Rb1) -- (Rt1) -- (Rt2) -- (Rt3) -- (Rt4) -- (Rt1);
      \draw[green,dashed] (Rb4) -- (Cbr);
      \draw[green,dashed] (Rt4) -- (Ctr);
    \end{tikzpicture}
  \end{subfigure}
  \caption{Above: Enlarged terms of the cross lattice, corresponding
  to hexagonal plaquettes and the adjoining square plaquettes. Below: pairs of
  adjacent terms, overlapping on square plaquettes.}
  \label{fig:cross}
\end{figure}
In considering the ``cross'' lattice we select subgraphs corresponding to
hexagonal plaquettes, together with the three adjoining square plaquettes.
In this case each such subgraph overlaps with three others, one for each
square plaquette. We find that $\eta=0.1997384500<\frac{1}{3}$.
\subsection{The honeycomb lattice with $n=1$ decoration}
\begin{figure} 
  \begin{subfigure}
    \centering
    \begin{tikzpicture}[scale=1.00, every node/.style={transform shape}]
      \foreach \i in {0,...,5}
      \foreach \j in {0,...,7} {
        \node[coordinate] (\i\j a0) at (4*sin{60}*\i,2*\j) {};
        \node[coordinate] (\i\j a1) at (4*sin{60}*\i+4/3*sin{60},2*\j) {};
        \node[coordinate] (\i\j b0) at (4*sin{60}*\i+2*sin{60},2*\j+2*cos{60}) {};
        \node[coordinate] (\i\j b1) at (4*sin{60}*\i+10/3*sin{60},2*\j+2*cos{60}) {};
        }
      \clip(11a0) rectangle (35b0);
      \foreach \i in {0,...,4}
      \foreach \j in {0,...,6} {
        \foreach \k in {a,b} {
          \foreach \l in {0,1} {
            \node[circle,fill=black,inner sep=0pt,minimum size=1mm] (\i\j\k\l p) at (\i\j\k\l) {}; }
          \foreach \a in {0,120,240} {
            \node[circle,fill=black,inner sep=0pt,minimum size=1mm] (\i\j\k;\a) at ($(\i\j\k 0)+(\a:2/3*sin{60})$) {}; }
        }
        \draw (\i\j a0) -- (\i\j a1);
        \draw (\i\j a1) -- (\i\j b0);
        \draw (\i\j b0) -- (\i\j b1);
        \pgfmathtruncatemacro\n{\i+1}
        \pgfmathtruncatemacro\m{\j+1}
        \draw (\i\j b0) -- (\i\m a1);
        \draw (\n\j a0) -- (\i\j b1);
        \draw (\i\j b1) -- (\n\m a0);
      }
      \foreach \i in {1,...,4}
      \foreach \j in {1,...,6} {
        \pgfmathtruncatemacro\ia{\i-1}
        \pgfmathtruncatemacro\ib{\i}
        \pgfmathtruncatemacro\ic{\i+1}
        \pgfmathtruncatemacro\ja{\j-1}
        \pgfmathtruncatemacro\jb{\j}
        \pgfmathtruncatemacro\jc{\j+1}
        \draw[cyan,dashed,very thick] plot [smooth cycle, tension=.4] coordinates {
          ($(\ia\ja b1)!.4!(\ia\ja b;0)$)
          ($(\ia\ja b1)!.4!(\ib\ja a;120)$)
          ($(\ib\ja b0)!.4!(\ib\ja b;240)$)
          ($(\ib\ja b0)!.4!(\ib\ja b;0)$)
          ($(\ib\jb b0)!.6!(\ib\jb b;240)$)
          ($(\ia\jb b1)!.6!(\ib\jb a;120)$)
        };
        \draw[magenta,dashed,very thick] plot [smooth cycle, tension=.4] coordinates {
          ($(\ia\jb a1)!.4!(\ia\jb a;0)$)
          ($(\ia\jb a1)!.4!(\ia\ja b;120)$)
          ($(\ib\jb a0)!.4!(\ib\jb a;240)$)
          ($(\ib\jb a0)!.4!(\ib\jb a;0)$)
          ($(\ib\jc a0)!.6!(\ib\jc a;240)$)
          ($(\ia\jc a1)!.6!(\ia\jb b;120)$)
        };
      }
    \node[below] at (current bounding box.south) {\fontsize{8}{8}\selectfont (b)};
    \end{tikzpicture}
  \end{subfigure}
  \begin{subfigure}
    \centering
    \begin{tikzpicture}[scale=1.00, every node/.style={transform shape}]
      \foreach \i in {0,...,5}
      \foreach \j in {0,...,7} {
        \node[coordinate] (\i\j a0) at (4*sin{60}*\i,2*\j) {};
        \node[coordinate] (\i\j a1) at (4*sin{60}*\i+4/3*sin{60},2*\j) {};
        \node[coordinate] (\i\j b0) at (4*sin{60}*\i+2*sin{60},2*\j+2*cos{60}) {};
        \node[coordinate] (\i\j b1) at (4*sin{60}*\i+10/3*sin{60},2*\j+2*cos{60}) {};
      }
      \node[Lv] (Llb) at (11a0) {};
      \node[Lv] (Lla) at ($(11a0)!.5!(01b1)$) {};
      \node[Lv] (Lld) at (00b1) {};
      \node[Lv] (Llc) at ($(Llb)!.5!(Lld)$) {};
      \node[Lv] (Lcb) at (11a1) {};
      \node[Lv] (Lm) at ($(Llb)!.5!(Lcb)$) {};
      \node[Lv] (Lca) at ($(11a1)!.5!(11b0)$) {};
      \node[Cv] (Cd) at (10b0) {};
      \node[Cv] (Cc) at ($(Lcb)!.5!(Cd)$) {};
      \node[Rv] (Rcf) at (10a1) {};
      \node[Rv] (Rce) at ($(Cd)!.5!(Rcf)$) {};
      \node[Rv] (Rrb) at (10b1) {};
      \node[Rv] (Rm) at ($(Cd)!.5!(Rrb)$) {};
      \node[Rv] (Rra) at ($(10b1)!.5!(21a0)$) {};
      \node[Rv] (Rrd) at (20a0) {};
      \node[Rv] (Rrc) at ($(Rrb)!.5!(Rrd)$) {};
      \draw[blue] (Lla) -- (Llb) -- (Llc) -- (Lld);
      \draw[blue] (Llb) -- (Lm) -- (Lcb) -- (Lca);
      \draw[blue,dashed] (Lcb) -- (Cc);
      \draw[red] (Cc) -- (Cd);
      \draw[green,dashed] (Rce) -- (Cd) -- (Rm);
      \draw[green] (Rce) -- (Rcf);
      \draw[green] (Rm) -- (Rrb);
      \draw[green] (Rra) -- (Rrb) -- (Rrc) -- (Rrd);
    \node[below] at (current bounding box.south) {\fontsize{8}{8}\selectfont (b)};
    \end{tikzpicture}
  \end{subfigure}
  \caption{Enlarged terms of the once-decorated honeycomb lattice}
  \label{fig:hexn1}
\end{figure}
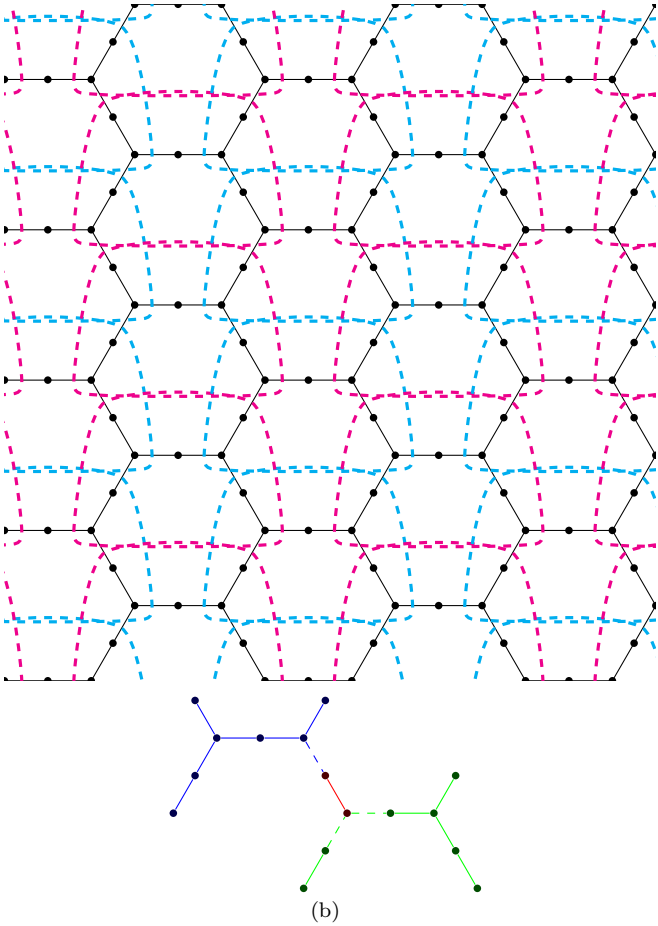
We now consider the honeycomb lattice with one decoration per edge. We
break rotational symmetry and consider each ``horizontal'' edge of the
original honeycomb lattice, connecting
vertices $v$ and $w$, together with the decorations on each of the edges
``above'' $v$ and $w$ and the full edge ``below'' each of $v$ and $w$, producing
the ``H'' shapes shown in Fig.~\ref{fig:hexn1}. Then each subgraph
overlaps with four others, at each of the four segments (half-edges) at its
boundaries. We find that $\eta=0.1530329085<\frac{1}{4}$.

In combination with the results of \cite{decorated1} and \cite{decorated2},
this means we can claim conclusively that the AKLT Hamiltonian on the
decorated honeycomb lattice is gapped for all $n > 0$.
\subsection{The square lattice with $n=1$ decoration}
\begin{figure} 
  \begin{subfigure}
    \centering
    \begin{tikzpicture}[scale=1.00, every node/.style={transform shape}]
      \foreach \i in {0,...,7}
      \foreach \j in {0,...,9} {
        \node[coordinate] (\i\j) at (2*\i,2*\j) {};
        }
      \clip(1,3) rectangle (9,11);
      \foreach \i in {0,...,6}
      \foreach \j in {0,...,8} {
        \node[circle,fill=black,inner sep=0pt,minimum size=1mm] (\i\j p) at (\i\j) {};
        \foreach \a in {0,90} {
          \node[circle,fill=black,inner sep=0pt,minimum size=1mm] (\i\j;\a) at ($(\i\j)+(\a:1)$) {}; }
        \pgfmathtruncatemacro\n{\i+1}
        \pgfmathtruncatemacro\m{\j+1}
        \draw (\i\j) -- (\i\m);
        \draw (\n\j) -- (\i\j);
      }
      \foreach \i in {1,...,3}
      \foreach \j in {1,...,4} {
        \pgfmathtruncatemacro\ia{2*\i-2}
        \pgfmathtruncatemacro\ib{2*\i-1}
        \pgfmathtruncatemacro\ic{2*\i}
        \pgfmathtruncatemacro\ja{2*\j-2}
        \pgfmathtruncatemacro\jb{2*\j-1}
        \pgfmathtruncatemacro\jc{2*\j}
        \draw[rotate=45,dashed,very thick,draw=cyan] ($(\ia\ja)!.5!(\ib\jb)$) ellipse (2.6 and 1.6);
        \draw[rotate=135,dashed,very thick,draw=magenta] ($(\ic\jb)!.5!(\ib\jc)$) ellipse (2.6 and 1.6);
      }
    \node[below] at (current bounding box.south) {\fontsize{8}{8}\selectfont (b)};
    \end{tikzpicture}
  \end{subfigure}
  \begin{subfigure}
    \centering
    \begin{tikzpicture}[scale=1.00, every node/.style={transform shape}]
      \node[Lv] (L0) at (2,2) {};
      \node[Lv] (L0a) at (3,2) {};
      \node[Lv] (L0b) at (2,3) {};
      \node[Lv] (L1) at (2,4) {};
      \node[Lv] (L1a) at (1,4) {};
      \node[Lv] (L1b) at (2,5) {};
      \node[Lv] (L2) at (4,4) {};
      \node[Lv] (L2b) at (3,4) {};
      \node[Lv] (L2a) at (4,3) {};
      \draw[blue] (L0a) -- (L0) -- (L0b) -- (L1) -- (L2b) -- (L2) -- (L2a);
      \draw[blue] (L1a) -- (L1) -- (L1b);
      \node[Cv] (C) at (4,2) {};
      \node[Cv] (Ca) at (4,1) {};
      \node[Cv] (Cb) at (5,2) {};
      \draw[blue,dashed] (L0a) -- (C) -- (L2a);
      \draw[red] (Ca) -- (C) -- (Cb);
      \node[Rv] (R0) at (4,0) {};
      \node[Rv] (R0a) at (3,0) {};
      \node[Rv] (R0b) at (4,-1) {};
      \node[Rv] (R1) at (6,0) {};
      \node[Rv] (R1a) at (5,0) {};
      \node[Rv] (R1b) at (6,1) {};
      \node[Rv] (R2) at (6,2) {};
      \node[Rv] (R2a) at (6,3) {};
      \node[Rv] (R2b) at (7,2) {};
      \draw[green] (R0) -- (R1) -- (R2);
      \draw[green,dashed] (Ca) -- (R0);
      \draw[green,dashed] (Cb) -- (R2);
      \draw[green] (R0a) -- (R0) -- (R0b);
      \draw[green] (R2a) -- (R2) -- (R2b);
    \node[below] at (current bounding box.south) {\fontsize{8}{8}\selectfont (b)};
    \end{tikzpicture}
  \end{subfigure}
  \caption{Enlarged terms of the once-decorated square lattice}
  \label{fig:squaren1}
\end{figure}
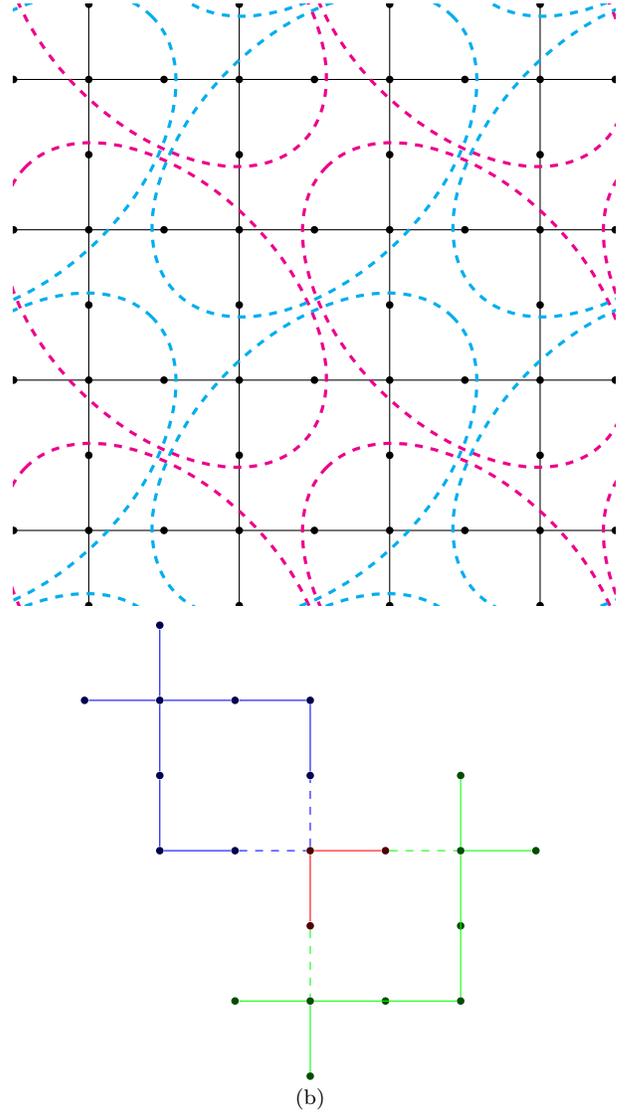
We conclude by considering the square lattice with one decoration per edge.
We first bipartition the dual lattice into sublattices $\mathcal{A}$ and
$\mathcal{B}$, and then further bipartition $\mathcal{A}$ into
$\mathcal{A}_1$ and $\mathcal{A_2}$. The subgraphs we will consider will
be of two types, represented in blue and red above in Fig.~\ref{fig:squaren1},
rotated relative to one another by $\frac{\pi}{2}$. The former consists of
a plaquette $p\in \mathcal{A}_1$ together with the two decorations
adjoining each of the upper-right and lower-left corners of $p$. The latter
consists of a plaquette $q\in \mathcal{A}_2$ together with the two decorations
adjoining each of the upper-left and lower-right corners of $q$. Each of
these subgraphs then have four 3-vertex intersections with other subgraphs, at
each of the four corners of the original plaquette. We find that
$\eta = 0.2203543174<\frac{1}{4}$.

In combination with the results of \cite{decorated2},
this means we can claim conclusively that the AKLT Hamiltonian on the
decorated square lattice is gapped for all $n > 0$.

\begin{table*}
  \[
  \begin{array}{l|l|l|c|c|c|c|c|c|c}
    &\text{Method}&\eta&\tilde{z}&\op{rank} E&\op{rank} E\wedge F&\dim\mathcal{H}_L&\dim\mathcal{H}_C&\dim\mathcal{H}_R&D_\text{tot}\\
    \hline\hline
    \text{Square-octagon}&I&0.1061446858&2&2^6&2^8&2^4&2^4&2^4&2^{12}\\
    &I&0.1589663310&4&&&&&&\\
    \hline
    \text{Star}&\text{I}&0.1110430220&4&2^4&2^5&2^3&2^3&2^3&2^9\\
    \hline
    \text{Cross}&\text{I}&0.1997384500&3&2^6&2^8&2^6&2^4&2^6&2^{16}\\
    \hline
    \text{Honeycomb}&\text{III}&0.1445124916&6&2^{12}&2^{18}&2^{10}&2^6&2^{10}&2^{26}\\
    \hline
    \text{Honeycomb, }n=1&\text{II}&0.1530329085&4&2^23^2&2^33^3&2^33&6&2^33^2&2^73^4\\
    \hline
    \text{Square, }n=1&\text{II}&0.2203543174&4&2^43^2&2^63^3&2^43^2&2^33&2^63&2^{13}3^4
  \end{array}
  \]
  \caption{For each of the configurations considered, the extracted value of
  $\eta$; the number of intersections $\tilde{z}$ of the type that this
  particular $\eta$ applies to;
  the rank of the projectors on the graphs $\Gamma_i$ and $\Gamma_j$,
  $\Gamma_i\cup\Gamma_j$,
  $L=\Gamma_i\setminus\Gamma_j$, $C=\Gamma_i\cap\Gamma_j$, and
  $R=\Gamma_j\setminus\Gamma_i$; and the total dimension
  $D_\text{tot} \equiv \dim(\mathcal{H}_L\otimes\mathcal{H}_C\otimes\mathcal{H}_R)$
  of the space on which the operators $E'+F'$ act.}
  \label{tab:data}
\end{table*}

\section{Completing the bound on the gap for the honeycomb lattice}
In order to produce a bound on the gap of the honeycomb-lattice AKLT lattice,
we must bound the original AKLT Hamiltonian relative to the altered Hamiltonian
whose gap we have bounded directly, by finding a bound $\gamma_0$ as in
\eqref{eq:originalbound}. Here $n_e$ will be 3 for each edge belonging to
one of the plaquettes in dual sublattice $\mathcal{B}$ and 1 otherwise.
Since the total physical dimension supported on one
of the chosen subgraphs, $4^{18}$, is far above our capacities, we instead
bound $\gamma_0$ using intermediate partitions of the subgraph $\Gamma$.
To make such an estimate we will need to project out degrees of freedom
much as we have already done; but here we will use the following
construction:
\begin{itemize}
  \item For a graph $\Gamma$, whose degrees of freedom factorize according
  to the vertices of $\Gamma$ as $\mathcal{H}=\bigotimes_{v\in\Gamma}\mathcal{H}_v$,
  \item Suppose that we have a frustration-free Hamiltonian defined as
  $H_0 = \sum_ic_iP_i$ ($c_i>0$),
  for a collection of projectors $P_i$ supported on 
  subgraphs $g_i\subset \Gamma$ (which cover $\Gamma$),
  with $\mathbbm{1}-\tilde{H}_0$ the projector
  onto the ground space of $H_0$
  \item Consider a collection of isometries
  $U_v:\mathcal{H}_v\to\mathcal{H}_v'$, corresponding to projectors
  $\Pi_v=U_vU_v^\dagger$, such that, for all $i$ such that $v\in g_i$, then
  $\Pi_v(\mathbbm{1}-P_i)=\mathbbm{1}-Pi$ (noting that, in the cases of
  interest, this will be an immediate consequence of the above Proposition)
  \item Let $H_1=\mathbb{P} H_0\mathbb{P}^\dagger$ and
  $\tilde{H}_1=\mathbb{P} \tilde{H}_0\mathbb{P^\dagger}$,
  where $\mathbb{P}=\bigotimes_{v\in\Gamma}U_v^\dagger$. Then
\end{itemize}
\begin{prop} If $H_1\geq \gamma_1\tilde{H}_1$, then for
\begin{equation}
  \gamma_0 = \min(\gamma_1,\min_ic_i),
\end{equation}
$H_0 \geq \gamma_0\tilde{H}_0$.
\end{prop}
Note that, by construction, each ``simplifying'' projector $\Pi_v$ commutes with
each of the projectors $P_i$ that comprise the Hamiltonian. Therefore, each
$\Pi_v$ commutes with $H_0$ as well as with each other. Thus, we can mutually
diagonalize these operators: for an eigenvector $\psi$ of $H_0$,
with eigenvalue $\lambda$
we can assume that $\psi$ is an eigenvector of each $\Pi_v$ as well. If there
is a $v$ such that $\Pi_v\psi=0$, then, for $g_i\ni v$,
$(\mathbbm{1}-P_i)\Pi_v\psi=0$ implies $P_i\psi = \psi$. In particular,
\begin{equation}
  \lambda = \langle\psi|H_0|\psi\rangle\geq c_i\geq \gamma_0.
\end{equation}
Now suppose that $\Pi_v\psi=\psi$ for all $v\in\Gamma$: in particular,
$\mathbb{P}^\dagger\mathbb{P}\psi=\psi$. Then $\mathbb{P}\psi\neq 0$ is
also an eigenvector of $H_1$ with eigenvalue $\lambda$. In particular,
if $\lambda\neq 0$ then $\lambda\geq \gamma_1$. $\square$

We first consider plaquette projectors. We can exactly diagonalize
$H_{p,0}$, the sum over the edges of the plaquette of the terms of the original
Hamiltonian; in doing so we determine that $H_{p,0} \geq \gamma_p H_p$
for $\gamma_p=0.3130520508$, with $\mathbbm{1}-H_p$ projecting out the ground
space of $H_{p,0}$. Then we can bound
\begin{equation}
  H_0 = \frac{1}{3}\sum_{\mathcal{B}\ni e\in \Gamma}H^{(e)} + \sum_{\mathcal{B}\not\ni e \in \Gamma}H^{(e)} \leq \frac{\gamma_p}{3}\sum_{\mathcal{B}\ni p\subset\Gamma}H_p + \sum_{\mathcal{B}\not\ni e \in \Gamma}H^{(e)}.
\end{equation}

We now consider pairs of plaquettes, as in the subgraphs $L$ and $R$ pictured
in Fig.~\ref{fig:honeycomb}, comprised of plaquettes $p$ and $q$ joined by
an edge $e$. Now
\begin{equation}
  H_0' = \frac{\gamma^{[p]}}{6}H_p + \frac{\gamma^{[p]}}{6}H_q+H^{(e)}.
\end{equation}
We will apply the above Proposition to these three terms, applied to a
compressed graph $\Gamma$ consisting of each of the two vertices of $e$,
the five remaining vertices of $p$ considered as one, and the five remaining
vertices of $q$ considered as one. Maximally projecting down the physical
spaces of these five-vertex groups, we determine a bound for $H_0'$ compared
with a combined two-plaquette projector $H_{p-q}$,
\begin{equation}
  \gamma^{[2p]} \leq \min(\gamma^{[2p]}_{1},\frac{\gamma^{[p]}}{6}) = 0.02571076873
\end{equation}
Thus we can write, for the three $\mathcal{B}$-plaquettes $\{p,q,r\}$,
\begin{equation}
  H_0 \geq \gamma^{[2p]}(H_{p-q} + H_{q-r} + H_{r-p}) \equiv \gamma^{[2p]}H_0''.
\end{equation}
Now we apply the above proposition again, to a three-vertex graph whose
vertices are the plaquettes $p$, $q$, and $r$, and obtain
\begin{equation}
  H_0'' \geq \gamma^{[4p]} \tilde{H},
\end{equation}
for $\gamma^{[4p]} = 0.7784203312$. Then we can write the overall bound on the
AKLT Hamiltonian as
\begin{equation}
  \gamma^{[2p]}\gamma^{[4p]}(1 - 6\eta) = 0.002660333395.\square
\end{equation}

\section{Note on precision and accuracy}
\label{sec:precision}
In general, we have relied on ARPACK methods when exact diagonalization has
been available; these generally afford us machine precision, which we confirm
by affirming $(E+F)\psi = (1\pm\eta)\psi$ to within less than $10^{-13}$. It is
for this reason that we report 10 digits of precision on our (admittedly
somewhat loose) bounds.

Due to time constraints, however, we have not employed full (machine) precision
when performing the second diagonalization step of Method III for the honeycomb
lattice, that is, finding the greatest eigenvalue of $O_2$ as defined by
\eqref{eq:iterative-ops}. (We note, for the sake of completeness, that we
apply the LM routine to this $O_2$, shifting it by a large constant in
order to ensure that the operator remains positive.) Having extracted the
greatest eigenvalue of $O_2$ with \texttt{tol} parameter $10^{-5}$, we
confirm that, for $\lambda$ the eigenvalue and $\psi$ the corresponding
eigenvector,
\begin{align}
  \left|(E+F)\psi-2\psi\right| &+ \left|(EF+FE)\psi-2\psi\right|\notag\\
  &+\left|\varepsilon\rho\psi-(3-\lambda)\psi\right| < 10^{-4}.
\end{align}
In particular, with $\lambda = 2.97451085252$, we can claim an error of no
more than $10^{-4}$, which keeps us well within the $\lambda<3$ that we need
to confirm that $E+F$ has no eigenvalues within $(\frac{3}{2},2)$.

\section{Concluding remarks}
We have closed the open cases in previous works on the spectral gap in
AKLT models in decorated honeycomb and square lattices. All these
decorated models have nonzero spectral gap, regardless of the number of
decorations $n>0$. More relevantly, we have proved the nonzero gap of
the original AKLT model on the honeycomb lattice, as well as those on
three other degree-3 Archimedean lattices. The lower bound on the
spectral gap for the honeycomb case is 0.002550333395. During the completion of
this manuscript, we
became aware of a recent preprint~\cite{LemmSandvikWang}, in which the gap of
the AKLT model on the honeycomb lattice is established, via a combination of
analytics and numerical DMRG methods. The spectral gap
in spin-2 AKLT models on the square lattice and other degree-4 lattices
should be possible to attack using our method. If the relevant effective
matrix $E'+F'$ to diagonalize is larger than the exact diagonalization
method or Lanczos, then one may need to resort to other numerical
methods. We further note that our method should apply in general to
models constructed using the so-called projected-entangled-pair-state
(PEPS) formalism, where the ground-space structure can be expressed in
terms of exact tensor networks and the parent Hamiltonian is usually
expressed in terms of projectors \cite{PerezGarciaPEPS}.

\begin{acknowledgments}
This work was partially supported by the National Science Foundation under
grant No. PHY 1915165. 
\end{acknowledgments}

\bibliography{bibs}

\end{document}